\documentclass[aps,prl,twocolumn,showpacs, superscriptaddress, floatfix]{revtex4}
\usepackage[dvipdfmx]{graphicx}

\usepackage{natbib}
\usepackage{amsmath}
\usepackage{amssymb}
\usepackage{bm}

\begin{document}

\title{Probing ultrafast spin dynamics through a magnon resonance in the antiferromagnetic multiferroic HoMnO$_3$}
\author{P. Bowlan}
\email{pambowlan@lanl.gov}
\affiliation{Center for Integrated Nanotechnologies, Los Alamos National Laboratory, Los Alamos, NM 87545, USA}
\author{S. A. Trugman}
\affiliation{Center for Integrated Nanotechnologies, Los Alamos National Laboratory, Los Alamos, NM 87545, USA}
\author{J. Bowlan}
\affiliation{Center for Integrated Nanotechnologies, Los Alamos National Laboratory, Los Alamos, NM 87545, USA}
\author{J.-X. Zhu}
\affiliation{Center for Integrated Nanotechnologies, Los Alamos National Laboratory, Los Alamos, NM 87545, USA}
\author{N. J. Hur}
\affiliation{Department of Physics, Inha University, Icheon 402-751, Korea}
\author{A. J. Taylor}
\affiliation{Center for Integrated Nanotechnologies, Los Alamos National Laboratory, Los Alamos, NM 87545, USA}
\author{D. A. Yarotski}
\affiliation{Center for Integrated Nanotechnologies, Los Alamos National Laboratory, Los Alamos, NM 87545, USA}
\author{R. P. Prasankumar}
\email{rpprasan@lanl.gov}
\affiliation{Center for Integrated Nanotechnologies, Los Alamos National Laboratory, Los Alamos, NM 87545, USA}

\date{\today}

\begin{abstract}
We demonstrate a new approach for directly tracking antiferromagnetic (AFM) spin dynamics by measuring ultrafast changes in a magnon resonance. We test this idea on the multiferroic HoMnO$_3$ by optically photoexciting electrons, after which changes in the spin order are probed with a THz pulse tuned to a magnon resonance. This reveals a photoinduced change in the magnon line shape that builds up over 5-12~picoseconds, which we show to be the spin-lattice thermalization time, indicating that electrons heat the spins via phonons. We compare our results to previous studies of spin-lattice thermalization in ferromagnetic (FM) manganites, giving insight into fundamental differences between the two systems. Our work sheds light on the microscopic mechanism governing spin-phonon interactions in AFMs and demonstrates a powerful approach for directly monitoring ultrafast spin dynamics.
\end{abstract}
\pacs{78.47.jh,75.50.Ee,75.85.+t} \maketitle


Ultrashort terahertz (THz) pulses are a powerful tool for probing or controlling material properties on a femtosecond time scale~\cite{kampfrath2013resonant,zhang2014dynamics}. Low energy resonances lying within the $\sim$0.3-3~THz bandwidth of typical THz pulses, such as phonons or spin waves (magnons), give access to specific material properties. For example, intense THz pulses have been used to control magnetism in antiferromagnets (AFM) by resonantly exciting magnons~\cite{kampfrath2011coherent,kubacka2014large, mukai2014antiferromagnetic}. Low energy resonances also offer a promising route for \textit{directly monitoring} the ultrafast evolution of specific material properties. This approach is especially useful for studying strongly correlated electron materials, such as multiferroics, which have coexisting and coupled order parameters (e.g., magnetism and ferroelectricity). Magnon resonances provide a unique way to directly probe spin order and its dynamics, particularly AFM spin order; this is more challenging than detecting ferromagnetic (FM) spin order, yet potentially very useful, since most multiferroic materials are AFM. This could also enable faster control and monitoring of spins, since spin dynamics are typically faster in AFMs than in FMs~\cite{kimel2004laser}.

Most previous time-resolved experiments on multiferroics involved photoexciting electrons with optical pulses and probing the resulting changes in the optical reflectivity~\cite{qi2012coexistence,sheu2012ultrafast,talbayev2015spin,jang2010strong,lee2013probing,shih2009magnetization}. Although transient optical reflectivity probes the temporal evolution of the photoexcited electron distribution, the measured time constants can still be sensitive to changes in magnetic order, since electronic and spin order are coupled. However, this is an $indirect$ probe of magnetic order that can be difficult to interpret, since the non-equilibrium electrons also couple to other degrees of freedom (e.g., phonons and other electrons). Time-resolved optical Kerr and Faraday rotation measurements more directly probe FM order~\cite{kimel2004laser}, and can be sensitive to the appearance of a net magnetization in optically excited AFM systems~\cite{Satoh2010NiO}. Second-order magneto-optical effects, such as magnetic linear birefringence and linear dichroism, are also sensitive to AFM order~\cite{Bossini2014AFM, Bossini2016} but the magnetic effects must be distinguished from other sources of birefringence or dichroism. Optical second harmonic generation (SHG) is another probe of AFM spin order~\cite{fiebig2002observation, duong2004ultrafast}; however, when applied to multiferroics it must be separated from the larger ferroelectric SHG signal. Finally, ultrafast AFM spin dynamics have been measured by resonant femtosecond x-ray diffraction using large scale free electron lasers or synchrotrons~\cite{tobey2012evolution,johnson2015magnetic,kubacka2014large,dean2016ultrafast}. These examples show that it is still not straightforward to directly probe spin dynamics in AFMs, which has limited the understanding of the governing mechanisms. For example, even thermal processes involving spins, although relatively well understood in FMs, show very different trends in AFM compounds that have not been extensively studied (compare, e.g. ~\cite{lee2013probing,talbayev2015spin,johnson2015magnetic} to \cite{averitt2002ultrafast}).

Here we directly monitor ultrafast AFM spin dynamics in antiferromagnetic HoMnO$_3$ after optical excitation by tracking changes in a magnon mode with a resonant THz pulse. Our measurements reveal a photoinduced change in the magnon line shape (amplitude, frequency and line width) occurring within $\sim$5--12~picoseconds (ps). Comparison of the line shape changes with simulations shows that the observed dynamics correspond to heating of the spin system through spin-phonon thermalization, as opposed to a direct transfer of energy from electrons to spins. To learn more about the microscopic origin of this heating, we compare our results, as well as measurements in similar AFM manganites~\cite{qi2012coexistence,talbayev2015spin,johnson2015magnetic}, to previous observations in FM manganites~\cite{averitt2001ultrafast}. While the microscopic mechanism governing spin-lattice thermalization in FM systems originates from spin-orbit coupling~\cite{yafet1963}, our study reveals very different trends in AFM systems, for which the underlying mechanism is not yet known. This sheds light on fundamental differences in spin-lattice interactions between FM and AFM systems, which may stem from the fact that a direct modulation of the exchange interaction by lattice vibrations can heat spins in an AFM, unlike in FMs~\cite{yafet1963}. Our measurements therefore give new insight into spin-lattice thermalization in AFMs, and most importantly, demonstrate a powerful experimental approach for directly monitoring specific properties, such as AFM order, in complex materials on an ultrafast timescale.

\begin{figure}[tb]
\includegraphics[width=3.2in]{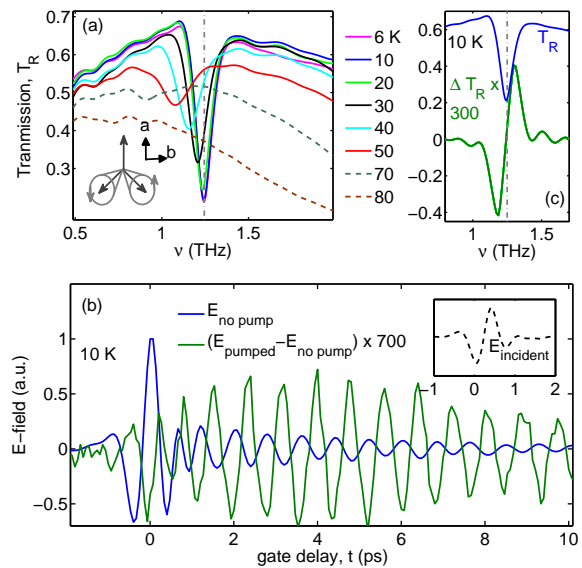}
\caption{ \label{fig:transmission} (Color online) (a) THz transmission spectrum, defined as $T_R (\nu)= |E_{no pump}(\nu)/E_{incident}(\nu)|$, where $E_{no pump}$ is the static THz spectrum transmitted through the HoMnO$_3$ crystal, and $E_{incident}$ is the THz spectrum before the crystal. The THz magnetic field is perpendicular to the $c$ axis of the crystal. The magnon mode that we excite is illustrated at the bottom left. (b) $E_{no pump}(t)$ at 10~K, and the differential transmitted field ($E_{pumped}(t)-E_{no pump}(t)$) for $F$=2.5~mJ/cm$^2$ and $\tau=$100~ps (green). The inset shows the incident THz pulse. (c) The differential transmission spectrum (green), $\Delta T_R(\nu) = |E_{pumped}(\nu)/E_{incident}(\nu)|-T_R(\nu)$, calculated from (b), and the 10~K static transmission spectrum from (a) (blue) for comparison.}
\end{figure}

HoMnO$_3$ consists of hexagonal layers of Mn$^{3+}$ ions at the centers of corner-sharing bipyramids of oxygen, with Ho$^{3+}$ ions located between these layers~\cite{rai2007spin}. Below 875~K, Ho-O displacements result in a ferroelectric polarization along the $c$-axis. Below the Ne\'el temperature, $T_N\sim$78~K, the Mn$^{3+}$ spins antiferromagnetically align in the $ab$-plane with 120$^\circ$ angles between them. At $T_{SR}=$~42~K there is a spin reorientation transition, where the Mn$^{3+}$ spins all rotate by 90$^\circ$ in the $ab$ plane~\cite{fiebig2002observation}. Magnetic order in HoMnO$_3$ leads to three different AFM $\vec q=0$ magnon modes emerging below $T_N$ ~\cite{penney1969far,vajk2005magnetic}. One of these involves precession of all of the spins about the $c$-axis~\cite{penney1969far}. The other two modes are degenerate and involve out-of-plane precessions of two of the spins about their equilibrium positions, as illustrated in the inset of Fig.~\ref{fig:transmission}(a).

The HoMnO$_3$ single crystal used in our experiments was grown in an optical floating zone furnace~\cite{hur2009giant}, with dimensions of 3$\times$3$\times$0.6~mm and the $c$-axis perpendicular to the polished surfaces. A more detailed description of our experimental setup is given in the Supplemental Material at [URL will be inserted by publisher]). Fig.~\ref{fig:transmission}(a) shows the THz transmission, $T_R$, as a function of temperature, $T$ , defined in the figure caption. At $T$=6~K an absorption peak centered around 1.25~THz is seen, corresponding to the magnetic dipole-active doubly degenerate magnon mode discussed above. As in similar compounds, the magnon peak broadens and shifts to lower frequencies as the temperature increases, since the sub-lattice magnetization decreases~\cite{talbayev2008magnetic, penney1969far,standard2012magnons,vajk2005magnetic}.

We photoexcited the HoMnO$_{3}$ crystal using 800~nm (1.55 eV) pulses with fluences of $F$=1--10~mJ/cm$^2$ (2.5~mJ/cm$^2$ corresponds to $\sim4\times10^{20}$ carriers/cm$^3$ at 10~K). Although the pump penetration depth is about 1000 times smaller than the THz probe penetration depth, our optical-pump, THz-probe (OPTP) signal comes only from the 200~nm thick pumped region. Comparison of our time scales to expected diffusion times ($>$20~nanoseconds) as well as all-optical pump-probe measurements on our crystal and optical-pump, x-ray diffraction probe measurements on a similar system~\cite{johnson2015magnetic}, where the penetration depths match, confirm this (see the Supplemental Material at [URL will be inserted by publisher] for more details). Absorption of our pump pulses excites \textit{on-site} Mn $d-d$ transitions~\cite{rai2007spin,souchkov2003exchange}. Fig.~\ref{fig:transmission}(b) shows the THz electric (E) field transmitted through the crystal (blue) at 10~K without photoexcitation as a function of the electro-optic sampling gate delay $t$. The long-lasting oscillations after the first few cycles originate from the magnon absorption at 1.25~THz, as seen by comparison with the input THz pulse shown in the inset. The photoinduced changes in the transmitted THz E-field at a pump-probe delay of $\tau=$100~ps  are shown in green, and in Fig.~\ref{fig:transmission}(c), we computed the photoinduced transmission change versus frequency ($\Delta T_R(\nu))$ (defined in the caption) from the curves in (b). Comparing this to the static transmission $T_R(\nu)$ from (a) reveals the most important feature of our data: the photoinduced changes occur \textit{primarily at the magnon resonance}. This is in stark contrast to most OPTP measurements, where free carrier dynamics dominate the photoinduced transmission changes, and demonstrates that in this material, THz pulses can be used to directly probe AFM spin dynamics. In particular, our measurements directly track the transfer of energy from the photoexcited electrons to this magnon mode, which we detect as changes in the line shape (i.e., amplitude, linewidth, and peak position).
	
\begin{figure}[tb]
\includegraphics[width=3.2in]{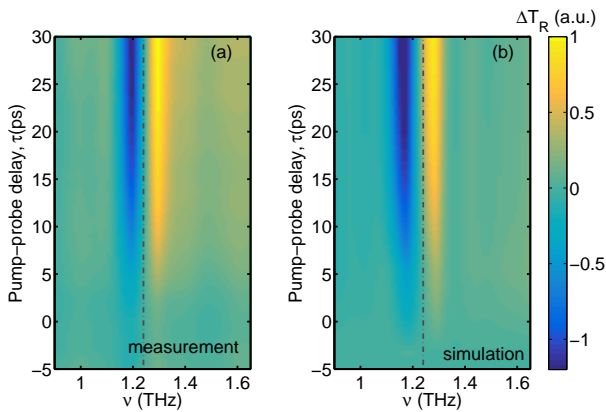}
\caption{ \label{fig:2D} (Color online) (a) The photoinduced transmission change $\Delta T_R(\nu,\tau)$ versus frequency and pump-probe delay.  (b) A simulation of the data in (a), calculated from the static THz transmission spectra in Fig.~\ref{fig:transmission}(a).}
\end{figure}

\begin{figure}[tb]
\begin{center}
\includegraphics[width=3.2in]{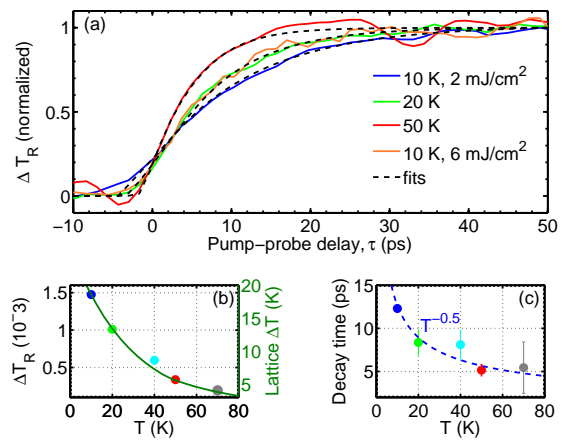}
\caption{ \label{fig:optp} (Color online) (a) Normalized optical-pump, THz-probe signals from HoMnO$_{3}$ at different temperatures and fluences. The dashed black lines are fits to the 2.5~mJ/cm$^{2}$ curves. (b) The amplitude of the signal versus temperature for $F$=2.5~mJ/cm$^{2}$, taken at the maximum of the OPTP signals. The green line shows the expected lattice heating on the right y-axis, calculated using a fluence of 2.5~mJ/cm$^2$ and the steady-state specific heat data from~\cite{zhou2006specific}. (c) The rise time of the pump-probe signals taken from the fits shown in (a), including a few additional temperature points. The blue dashed line is a fit to $constant\times T^{Q}$, where $Q$ = -0.5.}
\end{center}
\end{figure}

\begin{figure}[tb]
\begin{center}
\includegraphics[width=3.4in]{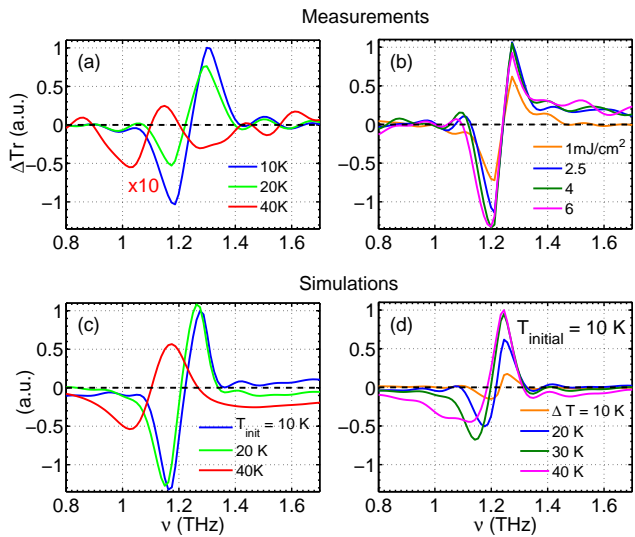}
\caption{ \label{fig:spectralchanges} (Color online) Measured photoinduced THz transmission change (defined in the caption of Fig.~\ref{fig:transmission}) at $\tau$=100 ps for different (a) sample temperatures with a pump fluence of 2.5~mJ/cm$^2$ and (b) for different pump fluences at 10~K. (c) Simulations of heating with a fluence of 2.5~mJ/cm$^2$ for different initial temperatures for comparison with (a). In these simulations, the final temperature, which varies with initial temperature, was taken from the green curve in Fig.~\ref{fig:optp}(b). (d) Simulations using $T_{initial}$=10~K and different final temperatures for comparison with (b). The simulations were done by taking the difference of the equilibrium transmitted E-fields with no optical-pump at different temperatures from Fig.~\ref{fig:transmission}(a). The spectra were all normalized to the value of the transmission change at the highest fluence for (a) and (c), and to the 10 K transmission change for (b) and (d).}
\end{center}
\end{figure}

To investigate this in more detail, we show the photoinduced transmission change versus frequency and pump-probe delay $\tau$ at 10~K, $\Delta T_R(\nu)$, as defined above (Fig.~\ref{fig:2D}(a)). This image shows the onset of the photoinduced magnon change along the pump-probe delay ($\tau$) axis. The spectral shape changes in the optical-pump, THz-probe (OPTP) signal with $\tau$ are very small, so that the time constant $\tau_r$ characterizing this process is approximately independent of the frequency or gate delay. Therefore, to determine its temperature dependence, we fixed the gate delay $t$ at a value which maximized the signal ($t$ = 4~ps for $T\leq$30 K and $t$=300~fs for higher temperatures, since the oscillations were more rapidly damped), and varied only $\tau$. The resulting $\Delta T_R(\tau)$ traces are shown in Fig.~\ref{fig:optp}(a); we note that after $\sim$50~ps, no significant changes were observed out to a delay of 500~ps. Fig.~\ref{fig:optp}(b) shows the amplitude of the OPTP signals versus temperature at $\tau$ = 100~ps, revealing that it decreases with increasing $T$ and is no longer detectable by $T_N$, further supporting the magnetic origin of the OPTP signal. We then extracted the time constant, $\tau_r$, shown in Fig.~\ref{fig:optp}(c), through a single exponential curve fit, revealing that it varies from $\sim$5-12~ps. Additional measurements of the fluence and temperature dependence of $\Delta Tr(\nu)$ at a fixed delay of $\tau$ = 100~ps are shown in Fig. \ref{fig:spectralchanges}(a)-(b).

We also note that our OPTP data exhibited an additional fast ($<$500~fs), spectrally broad feature, not only at the magnon resonance, but also spectrally overlapping with the transmitted THz pulse. This feature did not change significantly with $T$ and thus is unlikely to be magnetic in origin (see the Supplemental Material at [URL will be inserted by publisher]). This fast feature is not seen in Fig.~\ref{fig:optp}(a) because the larger magnon response dominates at the gate and pump-probe delays chosen here, and also because the time step in these measurements was 1.3~ps. Comparison to optical pump-probe measurements made on our HoMnO$_3$ crystals (see the Supplemental Material at [URL will be inserted by publisher]) and previous ultrafast measurements on manganites~\cite{averitt2002ultrafast, shih2009magnetization} and other AFM systems~\cite{Bossini2014AFM} allows us to attribute this fast process to electron-phonon thermalization.

The fact that phonon heating occurs before the observed $\sim$5-12~ps changes at the magnon resonance suggests that the latter process could be due to spin-lattice thermalization. If this is the case, the lattice temperature after gaining energy from the photoexcited electrons should match the final spin temperature, which we can estimate from our OPTP data. First, from the lattice specific heat \cite{zhou2006specific} and the pump fluence of 2.5~mJ/cm$^2$, we estimate the lattice heating ($\Delta T$) to be 18~K for an initial sample temperature of 10~K (see the right axis of Fig.~\ref{fig:optp}(b)). To compare this to the spin temperature, we performed simulations of the photoinduced changes in the magnon line shape versus delay (Fig.~\ref{fig:2D}(b)). This was done by first interpolating the data in Fig.~\ref{fig:transmission}(a) to get $T_R(\nu, T)$, and then calculating $\Delta T_R(\nu, \tau)$ from this matrix for an exponential rise in the spin temperature from 10~K to 28~K, using the 12~ps time constant from Fig.~\ref{fig:optp}(c). The good agreement between the simulation and measurements indicates that after the relaxation process shown in Fig.~\ref{fig:optp}, the spins are at approximately the same temperature as the lattice.

Additional simulations (Fig.~\ref{fig:spectralchanges}(c)-(d)) further demonstrate that the measured temperature and fluence dependence of the photoinduced changes in the magnon line shape are fully consistent with changes in the static transmission versus temperature. Fig.~\ref{fig:spectralchanges}(c) comes from varying $T_{initial}$ from 10~K to 40~K, given by the green curve in Fig.~\ref{fig:2D}(b). To simulate increasing the fluence (Fig.~\ref{fig:spectralchanges}(d)), we took differences in the measured static transmission spectra starting at $T_{initial}$ = 10~K and increasing $T_{final}$ from 20 to 50~K in steps of 10~K. The small discrepancies, especially at higher fluences, likely originate from our estimate of the spin and lattice temperatures from the heat capacity and fluence. In general, the magnon line width, frequency and amplitude all change with temperature in HoMnO$_3$ (Fig. 1(a)). While the photoinduced line shape changes are still clearly observable in Figs.~\ref{fig:2D} and \ref{fig:spectralchanges}(a)-(b), we note that these are quite small. For example, over the 18~K of heating that we observe in our OPTP measurements in Fig.\ref{fig:2D}, the magnon frequency changes only by about 30~GHz, close to the limit of our spectral resolution.

To understand the temperature dependence of the spin-lattice thermalization time $\tau_r$ (Fig. \ref{fig:optp}(c)), we begin by noting that a similar trend has been observed in several other AFM manganite compounds~\cite{talbayev2015spin,qi2012coexistence,johnson2015magnetic}; however, to the best of our knowledge, the reason for this behavior has not been definitively established. As a starting point, we consider the temperature dependence of spin-lattice relaxation in FM manganites, such as La$_{0.7}$Ca$_{0.3}$MnO$_3$, which is well described by a two-temperature model~\cite{averitt2002ultrafast}. Note that in this model, we only consider spins and phonons, since the electrons transfer energy to phonons on a much faster time scale than spin-lattice relaxation (as discussed above). Using this formalism, the spin-lattice thermalization time is given by $\tau_{sp} = {C_sC_p}/{(g(C_s+C_p))}$, where $C_s$ is the spin specific heat, $C_p$ is the lattice specific heat, and $g$ is the spin-phonon coupling constant. Because  $C_{p}>C_{s}$, this becomes $\tau_{sp} \approx {C_s}/{g}$. In ferromagnets, the temperature dependence of $\tau_{sp}$ follows that of $C_s(T)$ very closely, so that $g$ is deduced to be approximately temperature independent. In contrast, our measurements on HoMnO$_3$, as well as other ultrafast optical measurements on AFMs~\cite{talbayev2015spin,qi2012coexistence,johnson2015magnetic}, show that the spin-phonon thermalization time does not simply follow the spin specific heat, but also decreases monotonously with increasing temperature. We find that our data still fits to $\tau_{sp}(T) = C_{s}/g(T)$, if temperature dependence is incorporated into $g$. Fitting our data to a polynomial of the form $T^{Q}$ reveals that $Q$ = -0.5, as shown in Fig.~\ref{fig:optp}(c). Taking $C_{s} \propto T^2$, which is valid at the lower temperatures considered here \cite{vajk2005magnetic}, we find $g(T) = T^{-2.5}$.

The above comparison suggests that there are fundamental differences in the microscopic mechanisms governing spin-lattice relaxation in FM and AFM manganites. In FMs, hot phonons transfer energy to spins via spin-orbit coupling, as pointed out by Elliot and Yafet~\cite{yafet1963}. While this still occurs in AFMs, there is an additional term in the Hamiltonian that can moderate spin-lattice coupling, where lattice vibrations directly modulate the interatomic separation $r$, which can in turn modulate the exchange coupling $J(r)$. The total spin is conserved for this Hamiltonian, so such an interaction can never heat the spins in a FM, which requires reducing the total spin. This is however quite different in AFMs, since both the ground and high temperature excited states have zero spin. Therefore, spin conservation does not prevent lattice vibrations from heating the spins. Consequently, we suggest modeling spin-lattice relaxation in AFMs using a Hamiltonian which includes both this type of interaction, as in ref.~\cite{abrahams1952spin}, as well as spin-orbit coupling. Finally, we also point out that AFM compounds are typically insulators, while FMs are usually conductors, which probably also influences the observed differences in spin-lattice thermalization. We will address these issues further in a future publication, and have included more details in the Supplemental Material ([URL will be inserted by publisher]).

In conclusion, we demonstrated a powerful, table-top approach for directly probing ultrafast spin dynamics in AFMs, using a THz pulse resonant to a magnon absorption. This enables us to directly track the temporal evolution of the magnon line shape after photoexcitation in AFMs. To test this idea, we measured ultrafast photoinduced changes of the magnon line shape in the AFM multiferroic HoMnO$_3$. These changes were shown to be fully consistent with spin heating through thermalization with hot phonons on a $\sim$5-12~ps timescale. We find that the temperature dependence of spin-lattice relaxation in AFM HoMnO$_3$ is quite different from that previously observed in FM manganites, likely due to fundamental differences in the allowed pathways for energy transfer from phonons to spins. This provides new insight into spin-lattice thermalization in AFMs that should apply to many other systems. Even more generally, our measurements demonstrate a new way to track the dynamics of material properties by directly probing low energy excitations, such as magnons, electromagnons, and phonons, and thus offer a powerful approach for understanding the couplings between different degrees of freedom in complex materials.

\begin{acknowledgments}
This work was performed at the Center for Integrated Nanotechnologies, a U.S. Department of Energy, Office of Basic Energy Sciences user facility and also partially supported by the NNSA's Laboratory Directed Research and Development Program. Los Alamos National Laboratory, an affirmative action equal opportunity employer, is operated by Los Alamos National Security, LLC, for the National Nuclear Security Administration of the U. S. Department of Energy under contract DE-AC52-06NA25396. We
thank Rolando Vald\'es Aguilar for helpful discussions.
\end{acknowledgments}


\providecommand{\noopsort}[1]{}\providecommand{\singleletter}[1]{#1}%

\end{document}